\documentclass[twocolumn,amsmath,amssymb,prd,nofootinbib,showpacs]{revtex4}

\usepackage{graphicx}
\usepackage{dcolumn}
\usepackage{bm}
\usepackage{multirow}
\usepackage{epsf}
\usepackage{amsmath}

\def\be{\begin{equation}}
\def\ee{\end{equation}}

\def\bea{\begin{eqnarray}}
\def\eea{\end{eqnarray}}
\def\simgt{\stackrel{>}{{}_\sim}}
\def\simlt{\stackrel{<}{{}_\sim}}

\newcommand{\fNLl}{f_{\rm NL}^{\rm local}}
\newcommand{\fNL}{f_{\rm NL}^{}}
\newcommand{\LL}{{\mathrm{L}}}
\newcommand{\NL}{{\mathrm{NL}}}

\begin{document}

\title{Two statistical procedures for mapping large-angle non-Gaussianity}

\author{W. Cardona}
\affiliation{D\'{e}partement de Physique Th\'{e}orique,
Universit\'{e} de Gen\`{e}ve, 24 quai Ernest Ansermet,
CH-1211 Gen\`{e}ve 4, Switzerland}

\author{A. Bernui}
\affiliation{Observat\'orio Nacional \\
Rua General Jos\'e Cristino 77 \\
20921-400 Rio de Janeiro -- RJ, Brazil}

\author{M.J. Rebou\c{c}as}
\affiliation{Centro Brasileiro de Pesquisas F\'{\i}sicas\\
Rua Dr.\ Xavier Sigaud 150 \\
22290-180 Rio de Janeiro -- RJ, Brazil}

\date{\today}


\begin{abstract}
A convincing detection of primordial non-Gaussianity in the cosmic background radiation (CMB)
is essential to probe the physics of the early universe. Since a
single statistical estimator can hardly be suitable to detect the various possible
forms of non-Gaussianity, it is important to employ different statistical indicators
to study non-Gaussianity of CMB. This has motivated the proposal of a number
statistical tools, including two large-angle indicators based on skewness and
kurtosis of spherical caps of CMB sky-sphere. Although suitable to detect fairly
large non-Gaussianity they are unable to detect non-Gaussianity within the
Planck bounds, and exhibit power spectra with undesirable oscillation pattern.
Simulated  CMB maps are important tools to determine the strength, sensitivity
and limitations of such non-Gaussian estimators. Here we use several  thousands
simulated CMB maps to examine interrelated problems regarding  advances of
these spherical patches procedures. We examine whether a change in the choice
of the patches could enhance the sensitivity of the procedures  well enough to
detect large-angle non-Gaussianity  within the Planck bounds. To this end, a new
statistical procedure with non-overlapping cells is proposed and its capability is
established. We also study whether this new procedure is capable to smooth out
the undesirable oscillation pattern in the skewness and kurtosis power spectra of
the spherical caps procedure. We show that the new procedure solves this problem,
making clear this unexpected power spectra pattern does not have a physical origin,
but  rather  presumably arises from the overlapping obtained with the spherical caps
approach. Finally, we make a comparative analysis of this new statistical procedure
with the spherical caps routine,  determine their lower bounds for non-Gaussianity
detection, and make apparent their relative strength and sensitivity.
\end{abstract}

\maketitle

\section{Introduction} \label{Sec-1}

Recent cosmological observations are compatible with a nearly scale invariant spectrum of adiabatic perturbations~\cite{Planck1a,Planck1b}, which is predicted by a fair number of
slow-roll single scalar inflationary models. In this way,  there are many inflationary
models that fit this power-spectrum feature. This fact calls for new statistical procedures
to discriminate these models. A possible way to break this degeneracy among these models of the primordial universe is by studying deviation from Gaussianity of the cosmic microwave background (CMB)
temperature fluctuations~\cite{Bartolo04,Kawasaki2,Komatsu09,Komatsu10,Kawasaki3,Pogosyan,Sasaki,Chen}.

In single-field models of inflation, the amplitude of the local bispectrum can be written in terms of the slow-roll parameters and is undetectable ($ | \fNLl | \simlt 10^{-6} $)~\cite{Maldacena,Riotto1}.  On the other hand, both multi-field inflationary models and other alternative scenarios predict typically a detectable level of local non-Gaussianity \cite{NG-Alternative-Models}. Therefore,  a convincing   detection  of non-vanishing primordial non-Gaussianity of local type ($f_{\rm NL}^{\rm local} \gg 1 $) would rule out all the slow-roll single-field inflationary models and favour alternative models of the primordial universe~\cite{CreminelliZaldarriaga2004} (see also, e.g.,~\cite{Komatsu10} and references therein).

In the study of deviation from Gaussianity of the CMB temperature fluctuations data, one is particularly interested in the component coming from the early stage of the universe. However, it is well know that there exist several contributions which do not have a primordial origin. Some non-primordial contributions come from  unsubtracted diffuse foreground emission, unresolved point sources, possible systematic errors%
~\cite{Chiang03,Naselsky,Delabrouille,Aluri12,Cruz10,Saha}, and secondary anisotropies such as gravitational
weak lensing and the Sunyaev--Zel'dovich effect \cite{Aghanim08,Munshi,Pace,Novaes}.
Thus, the extraction of a possible primordial non-Gaussianity in CMB data is a challenging observational
and statistical endeavor.

In this context, on the one hand we have that  different statistical tools can potentially provide information about distinct features of non-Gaussianity ~\citep{Babich,Cabella10,Casaponsa11a,Casaponsa11b,Curto11,Donzelli, Ducout,Fergusson,Matsubara,NN,Noviello,Rossi,Pietrobon09,Pietrobon10a,Pratten,Smith,Vielva09,Yadav08,Zhao12}.
On the other hand, one does not expect that a single statistical estimator can be sensitive to all sorts of non-Gaussianity that may be present in observed CMB data. Thus, it is important to test CMB data for deviations from Gaussianity by using  different statistical tools. This has motivated a great deal of effort that has recently gone into the search for non-Gaussianity in CMB maps by employing several statistical
estimators and procedures%
~\cite{Abramo10,Bartolo10b,BTV,Bielewicz05,Bielewicz12,Copi10,Cruz07,Cruz09,Chiang07,Gruppuso10,%
Gruppuso11,Komatsu03a,Komatsu03b,Lew,McEwen,Park,Pietrobon09,Pietrobon10a,Rossi,Rossmanith,%
Raeth2,Raeth3,Vielva04,Vielva07,Wiaux,Yadav10}.

A simple way to  possibly detect deviation from Gaussianity in the data of a given CMB map is by
computing  the skewness and kurtosis from the whole set of values of the  temperature fluctuations.
This procedure would furnish  two dimensionless \textit{global} numbers for describing
the possible departures from non-Gaussianity of a CMB map. It is conceivable that in doing so
one would lose local (directional) information concerning the non-Gaussianity of the CMB
temperature fluctuations data. This has motivated the recent proposal of two large-angle
non-Gaussianity statistical indicators~\cite{BR2009,BR2010}, whose chief idea is the following:
starting from a given CMB temperature map one divides the CMB sphere $ \mathbb{S}^2 $ in $j$ (say)
spherical patches (subsets of CMB sky-sphere)  whose union covers the whole  CMB two-sphere.
Then one calculates the skewness ($S$) and kurtosis ($K$) of each spherical region and patches them
together to have two discrete functions $S(\theta,\phi)$ and $K(\theta,\phi))$ defined on $\mathbb{S}^2$.
These functions provide local measurements of the non-Gaussianity as a function of angular coordinates
$(\theta,\phi)$. Their Mollweide projections are skewness and kurtosis maps ($S-$map and $K-$map).
Motivated by the fact that simulated CMB maps with assigned type and amplitude of
non-Gaussianity are important tools to study the sensitivity, limitations, and
to determine the strength of non-Gaussian estimators, in a recent paper~\cite{BR2012},
by using overlapping spherical caps as the $j=1, \ldots ,N_\mathrm{c}$ regions
to calculate $S_j$ and $K_j$, it has been investigated  whether and to what extent these
non-Gaussian indicators have sensitivity to detect non-Gaussianity of local type,
particularly with amplitude within the  Wilkinson Microwave Anisotropy Probe (WMAP) bounds.
A systematic study was made by employing this statistical procedure to generate maps of skewness
and kurtosis from simulated maps equipped with non-Gaussianity of 
local type of various amplitudes. They have shown that $S$ and $K$ indicators, constructed
through spherical overlapping caps, can be used to detect large-angle local-type
non-Gaussianity only for reasonably large values of the non-linear parameter
$f_{\rm NL}^{\rm local}$, typically $ \fNLl \simgt 500 $. Thus, these indicators
have not enough sensitivity to detect deviation from Gaussianity of local type with the non-linear
parameter within the Planck bounds.
Moreover, they have found that the power spectra of all $S$ and $K$ generated maps exhibit an
unexpected \emph{zig-zag} pattern (lower values for even $\ell$ modes versus higher
values for neighboring odd modes), which does not seem to have a physical origin.
These unexpected oscillations in the values of the even and odd modes in the
power spectra could  have been induced by overlapping of the spherical caps
which contain  the temperature data used to construct the $S$ and $K$
maps.

Three possibly interrelated and pertinent questions arise naturally at this point.
First, whether a change in the choice of  the spherical patches (subsets of CMB sky-sphere)
in the above practical procedure to construct the functions $S(\theta, \phi)$  and $K(\theta, \phi)$
would enhance the sensitivity of constructive routine well enough to detect
non-Gaussianity of local type with $f_{\rm NL}^{\rm local}$ within the
Planck bounds. Second, whether a suitable non-overlapping choice of
spherical patches in the construction of $S(\theta, \phi)$  and $K(\theta, \phi)$
maps would be enough to smooth out the undesirable oscillation pattern in
the power spectra of the $S$ and $K$ maps.
Third, how could one compare the strength and weakness ot the new spherical
cells procedure with the  caps routine of ~\cite{BR2009,BR2010,BR2012} in
the attempts to detect large-angle ($\ell =1, 2, 3, 4$) non-Gaussianity.

Our primary aim in this paper is to address these questions by using several
thousands of simulated maps equipped with non-Gaussianity of local type of various
amplitudes, extending therefore the results of Refs.~\cite{BR2009,BR2010,BR2012} for a new
improved version of non-Gaussianity statistical indicators $S$ and $K$, where
the spherical patches are a complete set of appropriately chosen
equal area \emph{non-overlapping large pixels} or \textit{cells}  of the CMB two-sphere.
We examine the potentiality of the new version of the $S$ and $K$ statistical
non-Gausianity indicators and determine their efficacy and sensitivity to
detect non-Gaussianity of local type whose amplitude are within and above the
Planck limits.  We show that the new  non-Gaussianity indicators enhance
sensitivity of the constructive routine to  detect non-Gaussianity of local type
for $f_{\rm NL}^{\rm local} \simgt 500 $, but not for $f_{\rm NL}^{\rm local}$
within the Planck bounds. We also demonstrate that
the new improved version of the  indicators with non-overlapping patches do not
give rise to oscillations in the values of the even and odd modes in the
power spectra of $S$ and $K$ maps, solving this  undesirable
features of the previous indicators, and making  clear that the oscillations
do not have physical origin.

\section{Non--Gaussianity of Local Type and Simulated Maps}\label{Sec-2}

\subsection{Primordial non--Gaussianity of Local Type}

\begin{figure*}[th!]
\begin{center}
\includegraphics[scale=0.25]{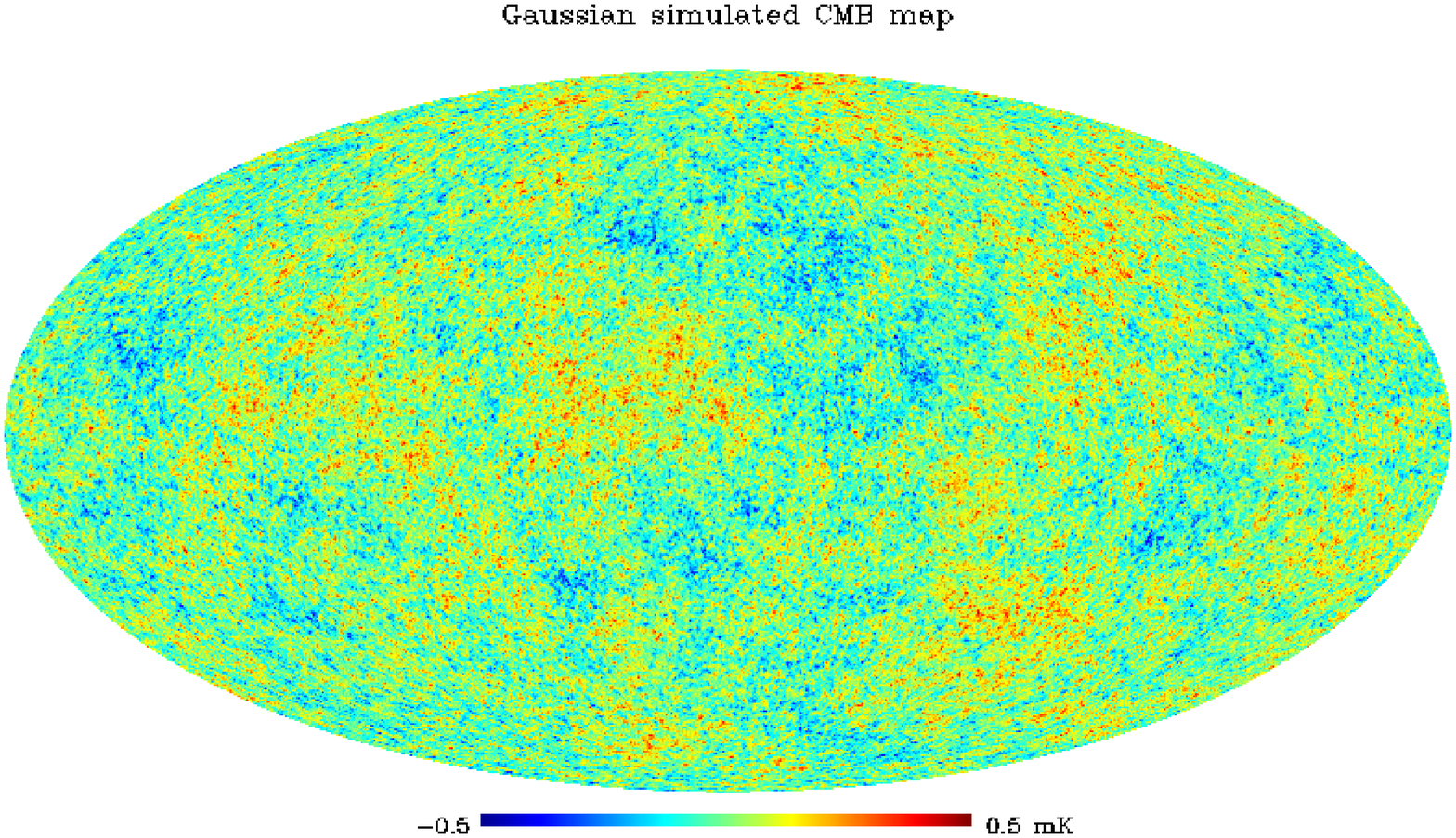}  
\hspace{8mm}
\includegraphics[scale=0.25]{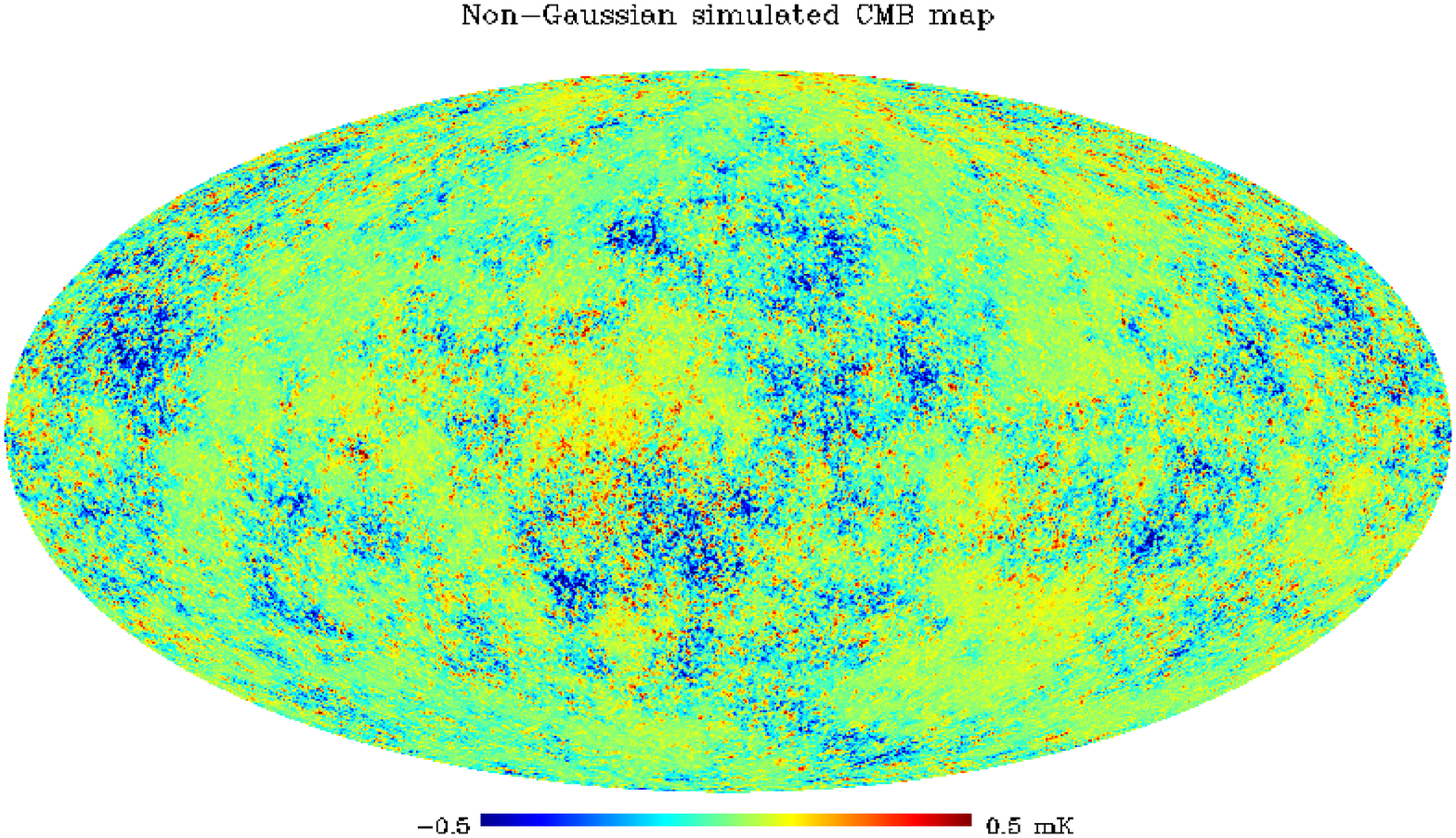}  
\caption{Gaussian (left) and non-Gaussian (right) simulated CMB maps generated for $\fNLl = 0$
and $ \fNLl = 5\,000 $, respectively.}
\label{cmbmaps}
\end{center}
\end{figure*}

Quantum fluctuations in the very early universe seeded primordial gravitational curvature perturbations
$\zeta(\mathbf{x},t)$ which later generated the non-homogeneities in the distribution of matter we
observe nowadays. In linear perturbation theory, the gauge invariant curvature perturbation $\zeta$
is related to the Bardeen's potential $\Phi(\mathbf{x},t)$ in the matter dominated era by
$\zeta= (5/3)\,\Phi$~\cite{Bardeen1980,Kodama1984}. On the other hand, on large scales the
Sachs-Wolfe effect~\cite{SachWolfe1967}  relates CMB temperature fluctuations and gauge
invariant curvature perturbations through
\be
\frac{\Delta T}{T} =  - \frac{\zeta}{5} = - \frac{\Phi}{3}\,,
\label{Sachs--Wolfe limit}
\ee
which allows to test inflationary models by comparing statistical properties of CMB data with
those of $\zeta$ (and thus $\Phi$) predicted by early universe models. Deviation from Gaussianity
can be studied by using the three-point correlation function of the curvature perturbations
$\Phi(\mathbf{x})$ or its Fourier transform $\Phi(\mathbf{k})$. The three-point correlation function
in Fourier space or bispectrum is
\begin{equation}
\langle \Phi(\mathbf{k_1}) \Phi(\mathbf{k_2}) \Phi(\mathbf{k_3})
\rangle = \delta^{(3)}(\mathbf{k_1 + k_2 + k_3})\,B^{}_\Phi(k_1,k_2,k_3) \,.
\label{Bispectrum definition 1}
\end{equation}
In the exam of primordial non-Gaussianity, the bispectrum of curvature pertubations
is rewritten in the form
\begin{equation}
B^{}_\Phi (k_1, k_2,k_3) = (2\pi)^3  \fNL \, F (k_1,k_2,k_3)\, ,
\label{Bispectrum definition 2}
\end{equation}
where $ \fNL $ is a dimensionless amplitude parameter which can be constrained by CMB
observations, and $F(k_1,k_2,k_3)$ is a function of the magnitude of the wave
numbers $(k_1,k_2,k_3)$ called shape of the bispectrum. Single field models of
inflation with standard kinetic terms and a Bunch-Davies vacuum  predict a maximum
for $ F (k_1,k_2,k_3) $ in the so-called local configuration $k_1 \approx k_2 \gg k_3$
and amplitude $ | \fNLl | \simlt 10^{-6} $.
This is a remarkable result because a convincing detection of local non-Gaussinity would
rule out all those single-field models~\cite{Maldacena,Riotto1,Komatsu09,Komatsu10}.

\subsection{Simulated CMB Maps}

Simulated CMB maps generated with a well-defined level and type of  non-Gaussianity are
essential tools to test the sensitivity and efficacy of non-Gaussian indicators.
Simulated CMB maps endowed with local non-Gaussianity can be generated by noting that
the Bardeen's potential $\Phi$ can reproduce the form of the bispectrum for the local
configuration when parametrized  in the form
\begin{equation}
\Phi (\mathbf{x})=\Phi_\LL (\,\mathbf{x})+\fNLl\,\left[ \,\Phi_\LL^2(\mathbf{x})
                            - \langle \,\Phi_\LL^2(\mathbf{x}) \,\rangle\,\right]\,,
\label{Non--Gaussianity of local type}
\end{equation}
with $ \Phi_\LL (\,\mathbf{x}) $ being a Gaussian field with zero mean. This parametrization was
used in Ref.~\cite{Elsner2009} to build an algorithm which generates simulated CMB temperature
and polarization maps endowed with local non-Gaussianity.  In this algorithm a simulated
map  with a fixed level of non-Gaussianity $\fNLl $ is defined by its spherical harmonic
coefficients
\begin{equation}
a_{\ell m} = a^\LL_{\, \ell m} + \fNLl \cdot a^\NL_{\,\, \ell m}\,,
\label{simulated maps with NG of local type}
\end{equation}
where $ a^\LL_{\, \ell m} $ and $ a^\NL_{\,\, \ell m} $ are, respectively, the linear and non-linear
spherical harmonic coefficients for the simulated CMB temperature map.%
\footnote{These linear and non-linear coefficients are available at
\href{http://planck.mpa-garching.mpg.de/cmb/fnl-simulations/}%
{http://planck.mpa-garching.mpg.de/cmb/fnl-simulations/}.}

Figure~\ref{cmbmaps} shows two examples of simulated CMB maps for $ \fNLl = 0 $ (Gaussian) and
$ \fNLl = 5\,000 $ (non-Gaussian) with grid resolution $ N_{side}=512 $, which we shall use throughout
this paper.

\section{Non-Gaussianity with spherical cells procedure} \label{Sec-3}

A simple way for describing deviation from  Gaussianity in CMB temperature fluctuations maps
is by calculating  the skewness $S=\mu_3/\sigma^3$, and the kurtosis $K=\mu_4/\sigma^4\!-\!3$
from the data, where $\mu_3$ and $\mu_4$ are the third and fourth central moments of the
distribution of the temperature anisotropies, and $\sigma$ is the variance.
Considering that $S$ and $K$ vanish for a Gaussian distribution, two statistical estimators
to measure large-angle deviation from Gaussianity in CMB maps were introduced in Ref.~\cite{BR2009},
from which a constructive general process can be formalized as follows.
Let $ \Omega_j \equiv \Omega(\theta_j, \phi_j) \in \mathbb{S}^2 $ be the set of points of each spherical
patch $j$ for $ j=1,\, \dots,\, N\, $. One can define functions $ S:\Omega_j \rightarrow \mathbb{R} $ and
$ K:\Omega_j \rightarrow \mathbb{R} $ , that assign to the $ j^{\rm th} $ spherical patch two
real numbers given by
\begin{eqnarray} 
S_j  & = & \frac{1}{N_{\mbox{\footnotesize p}} \,\sigma^3_j }
\sum_{i=1}^{N_{\mbox{\footnotesize p}}} \left(\, T_i\, - \overline{T} \,\right)^3 \;, \label{S_def}  \\
K_j  & = & \frac{1}{N_{\mbox{\footnotesize p}} \,\sigma^4_j }
\sum_{i=1}^{N_{\mbox{\footnotesize p}}} \left(\,  T_i\, - \overline{T} \,\right)^4 - 3 \;, \label{K_def}
\end{eqnarray}
where $T_i$ is the temperature at the $i^{\,\rm{th}}$ pixel, $\overline{T_j}$ is
the CMB mean temperature of the  $j^{\,\rm{th}}$ spherical patch  $ \Omega_j$,
$N_{\mbox{\footnotesize p}}$ is the number of pixels in  $ \Omega_j$,  
and  $\sigma^2 = (1/N_{\mbox{\footnotesize p}}^{}) \sum_{i=1}^{N_{\mbox{\footnotesize p}}^{}}
\left(\, T_i\, - \overline{T} \,\right)^2$ is the standard  deviation.
Now, the  set of all values $S_j$ and $K_j$ (calculated for all patches $\Omega_j$'s)
along with the angular coordinates of the center of the patch  $(\theta_j, \phi_j)$
are taken as measures of non-Gaussianity in the directions of the
center of each spherical patch  $\Omega_j$. This constructive process defines
two discrete functions $S=S(\theta,\phi)$ and $K=K(\theta,\phi)$ that give
the deviation from Gaussianity on the sky-sphere $\mathbb{S}^2$.

\begin{figure*}[tbh!]
\centering
\includegraphics[scale=0.25]{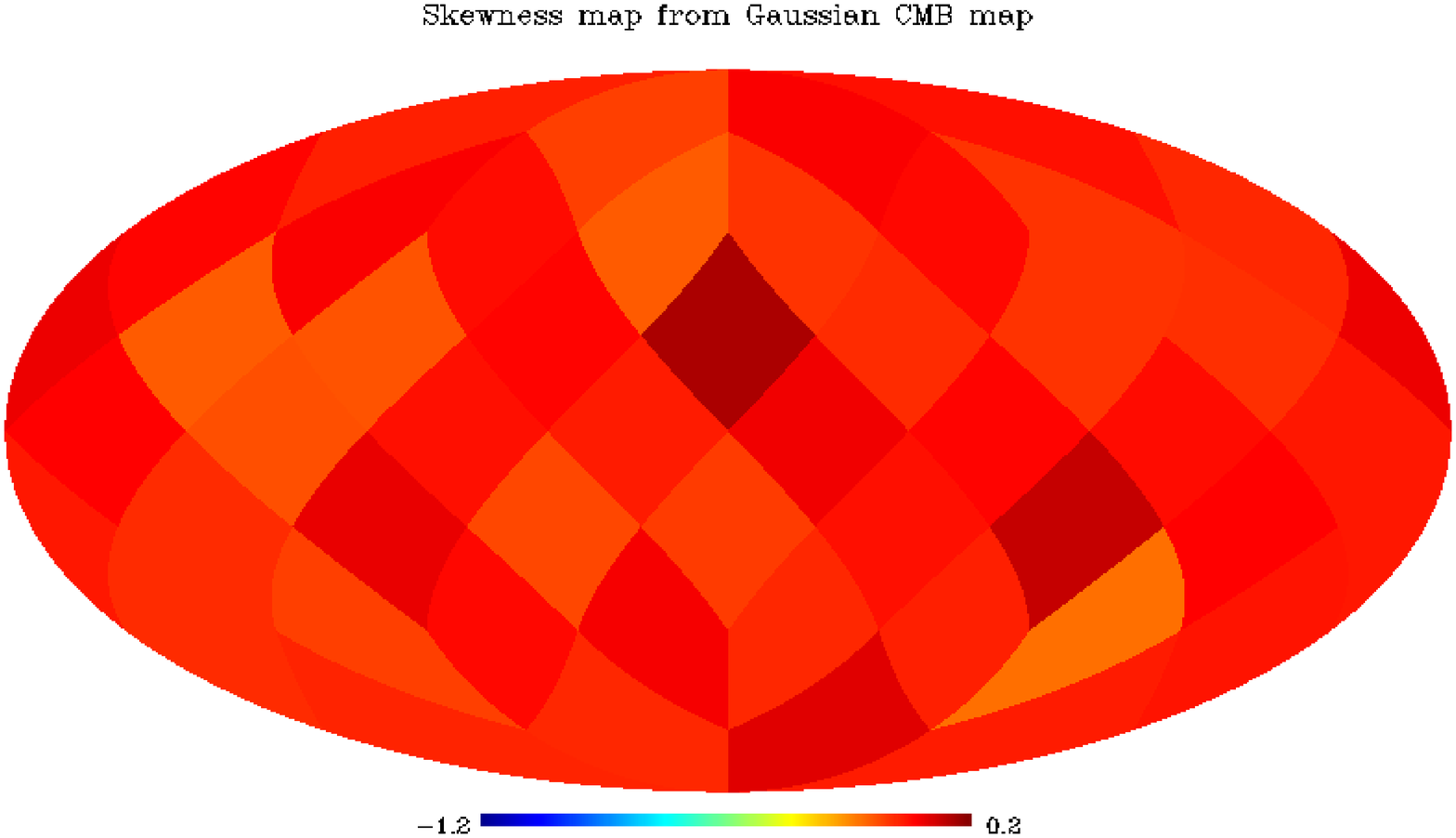}  
\hspace{8mm}
\includegraphics[scale=0.25]{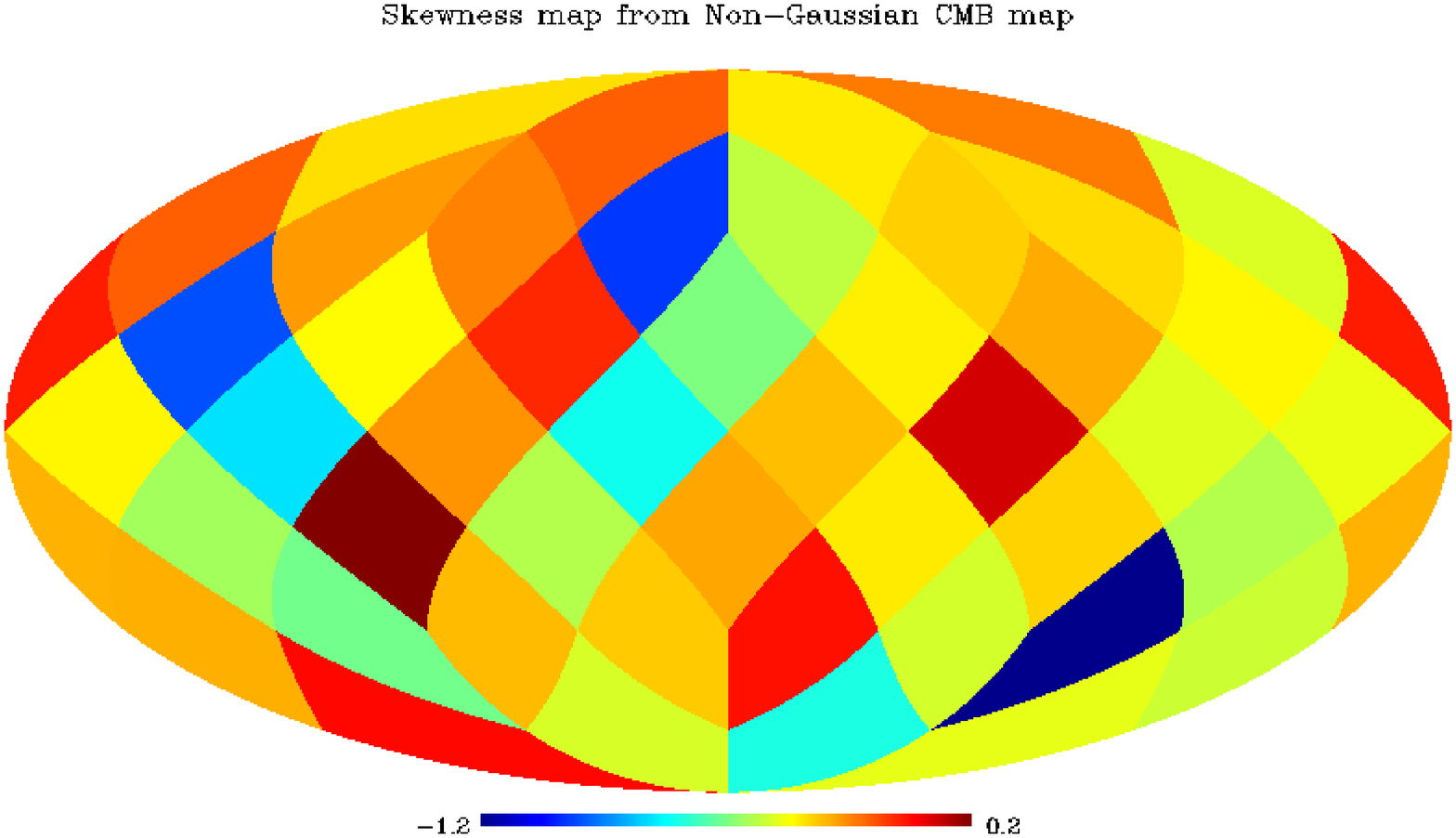}  
\caption{Skewness maps calculated by using the spherical cells procedure with $48$  cells along with
simulated  CMB input maps depicted in Fig.~\ref{cmbmaps}, whose amplitude of non-Gaussianity are
$\fNLl = 0$ (left panel) and $\fNLl = 5\,000$ (right panel), respectively. Each colored large
pixel (patch) is called a spherical cell.}
\label{skewness map}
\end{figure*}

\begin{figure*}[tbh!]
\begin{center}
\includegraphics[scale=0.25]{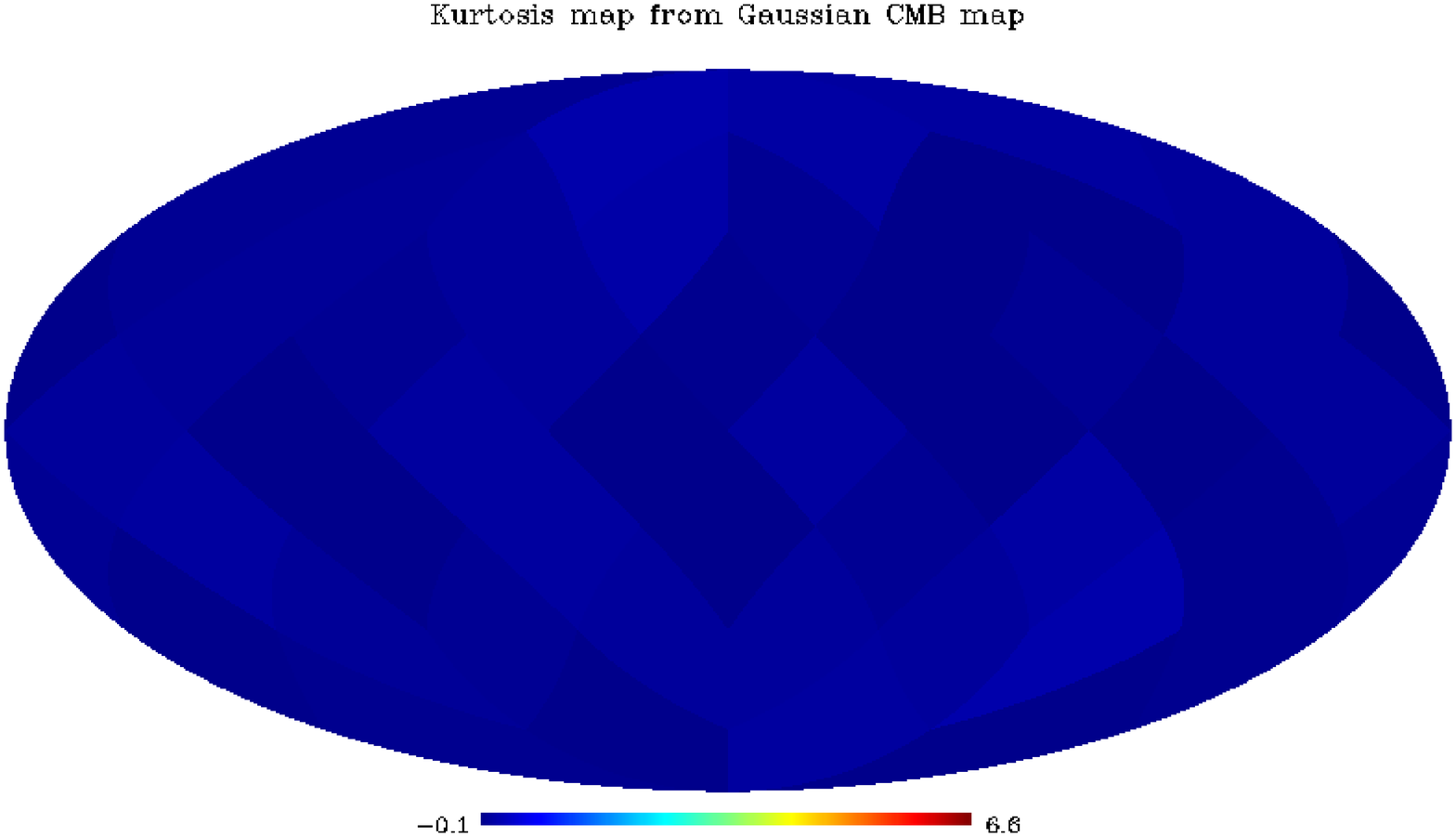}  
\hspace{8mm}
\includegraphics[scale=0.25]{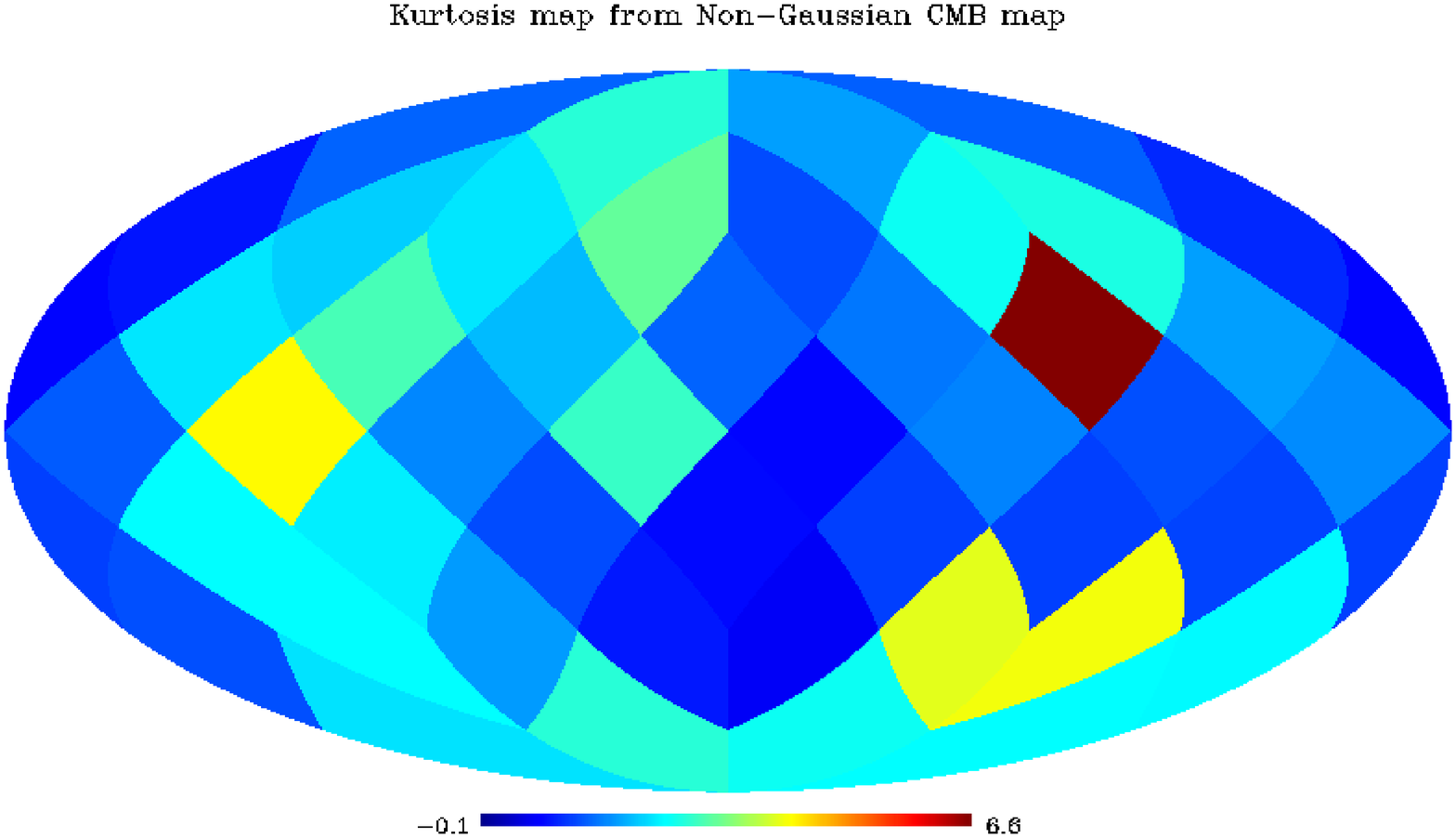}  
\caption{Kurtosis maps generated through the spherical cells routine with $48$  cells along with
simulated  CMB input maps depicted in Fig.~\ref{cmbmaps}, for which $\fNLl =0$ (left panel) and
$\fNLl = 5\,000$ (right panel), respectively.}
\label{kurtosis map}
\end{center}
\end{figure*}

In this work, to define  the skewness $S(\theta,\phi)$ and kurtosis $K(\theta,\phi)$
functions on $\mathbb{S}^2$, instead of overlapping spherical caps of
Refs.~\cite{BR2009,BR2010,BR2011,BR2012} we take as spherical patches appropriately chosen
equal area non-overlapping \textit{large} pixels generated by the
HEALPix partition of the CMB two-sphere~\cite{Gorski05} (see, e.g., Figure~\ref{skewness map}).
Throughout this article, we call \emph{spherical cell} or simply \textit{cell} each one of
these non-overlapping large pixels, and refer to the corresponding new routine
as spherical cells (or simply cell) procedure. In brief, the spherical cells procedure  \
can be formalized through the following steps:
\begin{enumerate}
\item
For a given CMB (input) map  with a determined grid resolution parameter $N_{\rm side}$,
divide the CMB sphere into $12$ equal area \textit{primary pixels} by using the
HEALPix~\cite{Gorski05} partition of the sphere;%
\footnote{With the HEALPix partition of the sphere, the total number of pixels for the
map equals $ N^{\rm total}_{\rm p} = 12\times N_{side}^2$.}
\item
Divide each one of the $12$ primary pixels in $N'^{\,2}_{\rm side} $ spherical cells,
producing therefore  $ N'^{}_{\rm p} = 12 \times N'^{\,2}_{\rm side}$ total number of
cells (see, e.g.,  Figure~\ref{skewness map}); 
\item
Use equations (\ref{S_def}) and (\ref{K_def}) to calculate $S_j(\theta_j,\phi_j) $
and \\
$K_j(\theta_j,\phi_j)$ for each $j=1,\cdots, N'^{}_{\rm p}$  cells;
\item
Patch together $S_j$ and $K_j$  to define the discrete functions $S(\theta,\phi)$ and
$K(\theta,\phi)$ on $\mathbb{S}^2$, whose Mollweide projection are, respectively, $S$ and $K$ maps.
These maps give a directional (geographical) distribution of values of skewness and kurtosis
calculated from a given CMB input map.
\end{enumerate}

As a practical application of this statistical procedure,
figures~\ref{skewness map} and~\ref{kurtosis map}  show, respectively, $S$ and $K$ maps with
$48$ spherical cells ($ N'_{\rm side}=2 $) computed from CMB input maps depicted in Fig.~\ref{cmbmaps}.
As expected, $S$ and $K$ maps for Gaussian  ( $\fNLl=0$)  CMB simulated  maps present a roughly uniform
low values (close colors) distribution for the skewness and kurtosis, whereas those maps calculated from the
non-Gaussian CMB simulated input maps with $\fNLl = 5\,000$ present inhomogeneous
higher values distribution for the skewness and kurtosis. 

Now, since the functions $S=S(\theta,\phi)$ and $K = K(\theta,\phi)$ are discrete
functions defined on $\mathbb{S}^2$ they can be expanded into their spherical harmonics,
and one can calculate their angular power spectra. Thus, for example, for  $K(\theta,\phi)$
one has
\begin{equation}
K(\theta,\phi) = \sum_{\ell=0}^\infty \sum_{m=-\ell}^{\ell}
b_{\ell m} \,Y_{\ell m} (\theta,\phi) \; ,
\end{equation}
and  the corresponding angular power spectrum
\begin{equation}
K_{\ell} = \frac{1}{2\ell+1} \sum_m |b_{\ell m}|^2 \; .
\end{equation}
Clearly, one can similarly expand the skewness function $S(\theta,\phi)$ and
calculate its angular power spectrum $S_{\ell}$.
In the next section we shall use the power spectra $S_\ell$ and $K_\ell$
to assess the departure from Gaussianity, i.e. to calculate the statistical
significance of such a deviation by comparison with the corresponding power spectra
calculated from input Gaussian maps ($f_{\rm NL}^{\rm local}=0$), and
determine their suitability of the cellular procedure to detect non-Gaussianity.

\section{Analyses and results} \label{Sec-4}

In this section we report the results of our analyses of non-Gaussianity carried out
by using the  $S(\theta,\phi)$ and $K(\theta,\phi)$ functions and associated
$S$ and $K$ maps, constructed through the spherical cells procedure, described
in Sec.~\ref{Sec-3}, along with several thousands of simulated CMB maps equipped
with non-Gaussinity of local type.

First, we note that in order to have a reference for comparing non-Gaussinity,  we  have
generated $1\,000 $ $S$ and $1\,000 $  $K$ maps from $1\,000 $  CMB Gaussian ($\fNLl=0$)
simulated maps, and have computed their angular power spectra $S^{\mathbf{i} }_{\,\ell}$
and $K^{\mathbf{i} }_{\,\ell}$ ($\,\mathbf{i} = 1,\,\,\cdots,1\,000\,$ is an enumeration index)
to have the mean angular power spectra $S_\ell = (1/1000) \sum_{\mathbf{i} = 1}^{1000}
S_{\,\ell}^{\mathbf{i}}\,$ and similarly the mean spectra $K_\ell $.
A similar procedure has been used several times to calculate the mean angular power spectra
from  $1\,000$ $S$ maps, and of $1\,000$  $K$ maps computed from $1\,000$ CMB
non-Gaussian maps with different $\fNLl$ in each instance. 

We have measured the statistical significance of the deviations from Gaussinity through
a $\chi^2$ test to find out the departure from goodness of the fit between mean power
spectra calculated from Gaussian maps ($f_{\rm NL}^{\rm local}=0$), $S_\ell^{G}$ and
$K_\ell^{G}$, and mean angular power spectra computed from non-Gaussian
($f_{\rm NL}^{\rm local} \neq 0$) maps, $ S_\ell^{NG} $ and $ K_\ell^{NG} $.
In other words,  we have used $\chi^2$ test along with the power spectra
to assess the significance of the deviation from Gaussianity.

In this way, for the indicator $S$  (similar expression holds for $K$) one
has
\begin{equation}
\chi^2_{S_\ell} = \frac{1}{n-1} \sum_{\ell=1}^{n}\frac{\left({S_\ell^{NG}} - {S_\ell^{G}}
\right)^2}{(\,{\sigma_\ell^G}\,)^2}\,, \label{chi squared}
\end{equation}
where $S_\ell^{G}$ and $S_\ell^{NG}$ are the mean values for each $\ell$ mode,
$(\sigma_{\ell}^{G})^2 $  is the variance calculate from Gaussian maps, and $n= 4$ is the
highest multipole permitted for the grid resolution of $48$ cells we have employed.

Clearly the greater is this value the smaller is the associated $\chi^2$ probability, i.e. the
smaller is the probability that the mean multipoles values $S_{\ell}$ and $K_{\ell}$ calculated
from a given non-Gaussianity map ($\fNLl \neq 0 $) and the mean multipole values obtained
from  ($\fNLl=0$) agree. We have employed  this  procedure to the analyses
whose results we shall report in the following.

\begin{table}[bth!]
\centering
\begin{tabular}{|c|c|c|c|c|}
\hline
\multicolumn{1}{|c|}{\multirow{2}{*}{$ \fNLl $}} & \multicolumn{2}{|c|}{$48$ cells} & \multicolumn{2}{|c|}{$ 192 $ cells} \\ \cline{2-5}
                   & $ \chi^2_{S_\ell} $         & $\chi^2_{K_\ell}$         & $\chi^2_{S_\ell}$          & $\chi^2_{K_\ell}$ \\ \cline{1-5}
$ 20 $         & $ 1.01\times10^{-5} $  & $ 2.41\times10^{-5} $ & $ 2.57\times10^{-6} $ & $ 3.46\times10^{-6} $  \\ \cline{1-5}
$ 56 $         & $ 1.04\times10^{-3} $  & $ 9.43\times10^{-4} $ & $ 2.42\times10^{-4} $ & $ 1.53\times10^{-4} $   \\ \cline{1-5}
$ 100 $       & $ 9.62\times10^{-3} $  & $ 6.63\times10^{-3} $ & $ 2.22\times10^{-3} $ &  $ 1.13\times10^{-3}  $  \\ \cline{1-5}
$ 400 $       & $ 3.14\times10^{-1} $    &  $ 1.80\times10^{-1} $&  $5.63\times10^{-1}$ &  $ 2.60\times10^{-1} $  \\ \cline{1-5}
$ 500 $       & $ 5.88 $                        & $ 4.05 $                       & $ 1.38 $                      & $ 7.04\times10^{-1} $ \\ \cline{1-5}
$ 1000 $     & $ 9.19\times10 $          & $ 1.74\times10^2 $     & $ 2.33\times10 $        & $ 4.04\times10 $ \\ \cline{1-5}
$ 3000 $     & $ 3.02\times10^3 $      & $ 2.67\times10^5 $     & $ 1.13\times10^3 $    & $ 1.56\times10^5 $ \\ \cline{1-5}
$ 5000 $     & $ 7.76\times10^3 $      & $ 4.95\times10^5 $     & $ 4.10\times10^3 $    & $ 2.02\times10^{5} $ \\ \cline{1-5}
\hline
\end{tabular}
\caption{$\chi^2$ values calculated from the mean power spectra $S_\ell$ and $K_\ell$ of $S$ and $K$ maps
computed from simulated temperature maps for $8$ values of $\fNLl$ including the Planck limits. The
spherical cells procedure with $48$ and $192$ cells was employed to generate the $S$ and $K$ maps}
\label{Table-1}
\end{table}

\begin{figure*}[tbh!]
\begin{center}
\includegraphics[scale=0.45]{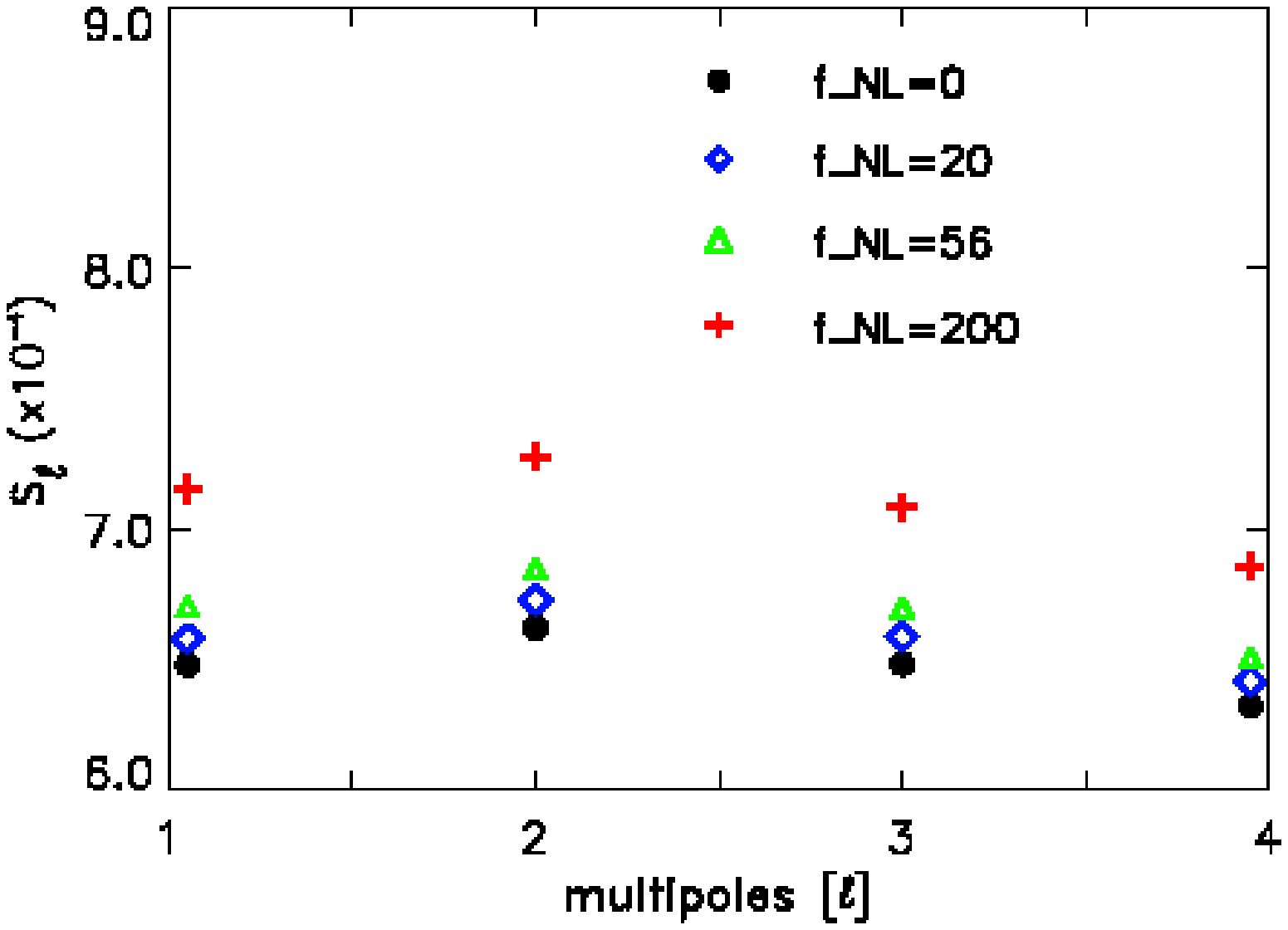}   
\hspace{8mm}
\includegraphics[scale=0.45]{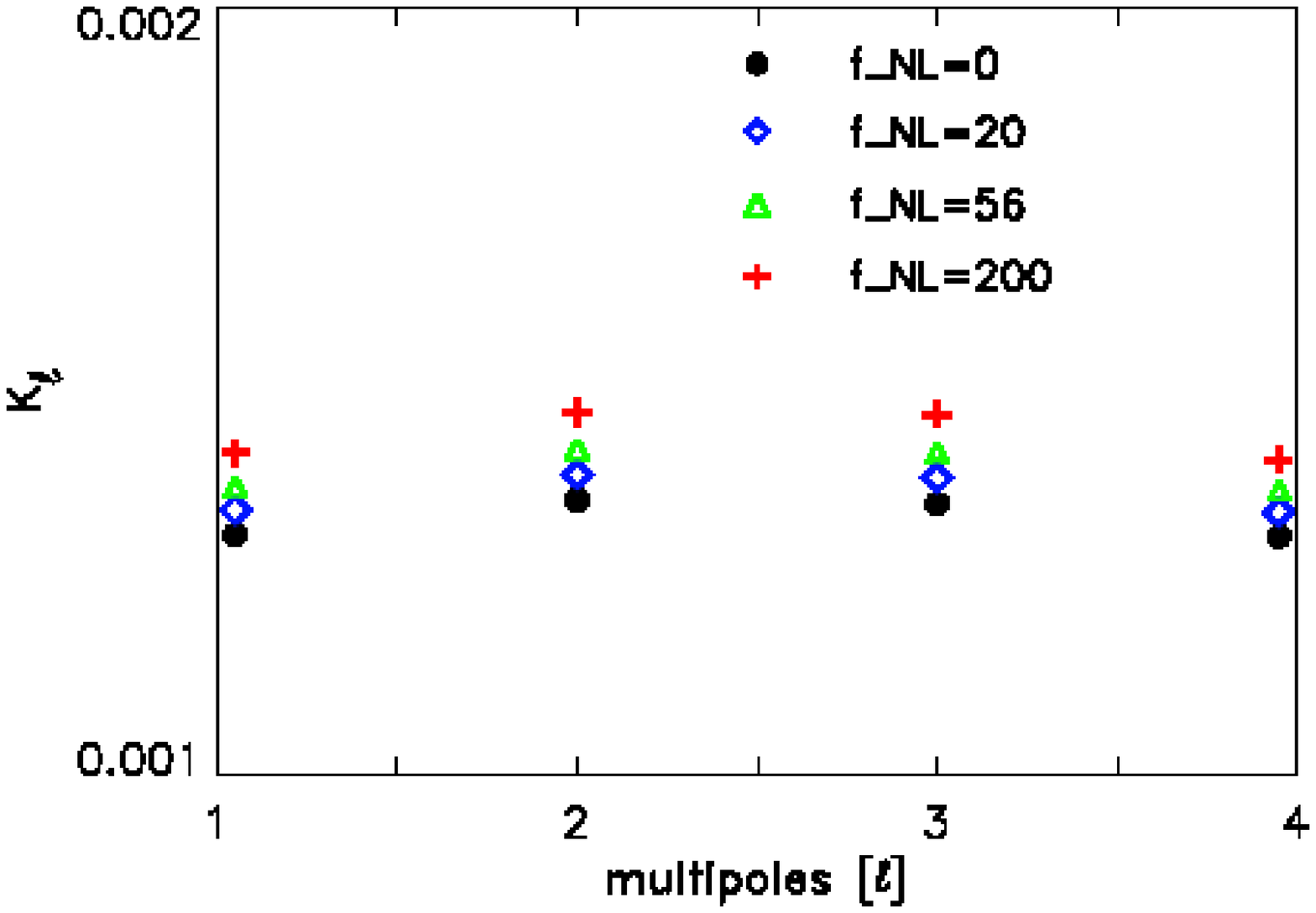}   
\caption{Low $\ell$ angular power spectra $S_\ell $ (left panel) and $K_\ell $ (right panel) of $S$ and
$K$ maps generated from simulated CMB input maps equipped with non-Gaussianity of local type whose values
of the amplitude parameter are $\fNLl = 20, 56$ (Planck limits) as well as $\fNLl = 0$ (Gaussian)
and  $\fNLl=200$ (illustrative value).}
\label{Fig-4}
\end{center}
\end{figure*}

In order to choose the most suitable number of cells in the procedure, we have examined
how the behavior of  $S$ and $K$  indicators is affected by the number of 
cells used to construct $S(\theta,\phi)$ and $K(\theta,\phi)$ functions and associated maps.
To this end, we have used  non-Gaussian simulated CMB maps with local non-Gaussian
amplitude  $\fNLl$ in the limits of the interval found by Planck experiment,%
\footnote{We stress that given the resolution of the simulated maps used in this work,
we have taken the $\fNLl$ limits for $\ell_{max}=500$ of the SMICA cleaned maps as reported
in the table $16$ of Ref.~\cite{Planck-NG}, that is, $ 38\pm18 $.}
as well as several other values for the amplitude of local non-Gaussinity (see Table~\ref{Table-1}).
For each value of $\fNLl$ we have generated $1\,000$ simulated CMB input maps, from which we  have
calculated $1\,000$ $S$ and $1\,000$ $K$ maps by using $48$ and $192$ cells and the associated
power spectra $S_{\,\ell}^{\mathbf{i}}\,$ and $K_{\,\ell}^{\mathbf{i}}\,$
to have the mean power spectra $S_\ell^{NG}$ and $K_\ell^{NG}$, which were
used to finally compute the $\chi^2_{S_\ell}$ and $\chi^2_{K_\ell}$ collected together
in Table~\ref{Table-1}.%
\footnote{To construct the Table \ref{Table-1}  we have generated a total of
$45\,000$, of which $9\,000$ were simulated temperature CMB (input) maps for
different $\fNLl$ including the Gaussian maps, and $18\,000$ $S-$maps and
$18\,000$ $K-$maps computed in each case by using both $48$ and $192$ cells.}
The results of the $ \chi^2 $ analysis in this table show that
the spherical cells  procedure, with both $48$ and $192$ cells, do not have sufficient
sensitivity to detect deviation from non-Gaussianity for the values of the non-Gaussian
parameter within the Planck bounds $ 20 \leq  f_{\rm NL}^{\rm local} \leq 56$. Indeed, for
simulated maps whose amplitude parameter $\fNLl$ are equal to the Planck bounds one
has a negligible value of $\chi^2$, which makes apparent that
there is no significant overall departure of the mean power spectra
$S_\ell^{NG}$ and $K_\ell^{NG}$ and the associated power spectra obtained from
Gaussian maps ($\chi^2-$probability is virtually equal to 1).
In  Fig.~\ref{Fig-4} we illustrate this result by showing the nearly
overlapping  of the symbols corresponding to $\fNLl=0, 20, 56$. For
visual comparison, in Fig.~\ref{Fig-4} we also present the result for $\fNLl=200$.

To the extent that the average $S_{\ell}$ and $K_{\ell}$ obtained
from simulated CMB maps endowed with $f_{\rm NL}^{\rm local}= 500$
are within $95\%$ average values of $S_{\ell}$ and $K_{\ell}$ for $f_{\rm NL}^{\rm local}= 0$,
Table~\ref{Table-1} shows that spherical cells procedures are not suitable to detect primordial
non-Gaussianity of local type in CMB maps smaller than $\fNLl=500$.
Table~\ref{Table-1} also shows that for non-Gaussinity amplitude $\fNLl \geq 500$
the spherical cells procedure with $48$ cells is more suitable than with $192$ cells,
inasmuch as it detects  values for $\chi^2$ greater than the values obtained for $192$ cells.
This favors the use of $48$ cells in the comparative analysis of the spherical cells and
caps procedures, which we shall discuss in what follows.  

\begin{table}[tbh!]
\centering
\begin{tabular}{|c|c|c|c|c|}
\hline
\multicolumn{1}{|c|}{\multirow{2}{*}{$ \fNLl $}} & \multicolumn{2}{|c|}{$48$ cells} & \multicolumn{2}{|c|}{$ 3072 $ Caps} \\ \cline{2-5}
                   & $ \chi^2_{S_\ell} $         & $\chi^2_{K_\ell}$         & $\chi^2_{S_\ell}$          & $\chi^2_{K_\ell}$ \\ \cline{1-5}
$ 20 $         & $ 1.01\times10^{-5} $  & $ 2.41\times10^{-5} $ & $ 8.33\times10^{-3} $ & $ 5.07\times10^{-3} $  \\ \cline{1-5}
$ 56 $         & $ 1.04\times10^{-3} $  & $ 9.43\times10^{-4} $ & $ 8.58\times10^{-3} $ & $ 5.3\times10^{-3} $   \\ \cline{1-5}
$ 500 $       & $ 5.88 $                        & $ 4.05 $                       & $ 4.5\times10^{-4} $   &  $ 3.58\times 10^{-3} $\\ \cline{1-5}
$ 1000 $     & $ 9.19\times10 $          & $ 1.74\times10^2 $     & $ 6.05\times10^{-3} $ & $ 9.79\times10^{-2} $  \\ \cline{1-5}
$ 3000 $     & $ 3.02\times10^3 $      & $ 2.67\times10^5 $     & $ 2.16\times10^{-1} $ & $ 2.77\times10 $ \\ \cline{1-5}
$ 5000 $     & $ 7.76\times10^3 $      & $ 4.95\times10^5 $     & $ 5.64\times10^{-1} $ & $ 1.53\times 10^{2} $  \\ \cline{1-5}
\hline
\end{tabular}
\caption{$\chi^2$ values calculated from the mean power spectra $S_\ell$ and $K_\ell$ of $S$ and $K$ maps
computed from simulated temperature maps for $6$ values of $\fNLl$ including the Planck limits. The
the spherical cells procedure with $48$ and the spherical caps method with $3\,072 $ caps were used for
comparison of these two approaches.}
\label{Table-2}
\end{table}

\begin{figure*}[tbh!]
\centering
\includegraphics[scale=0.45]{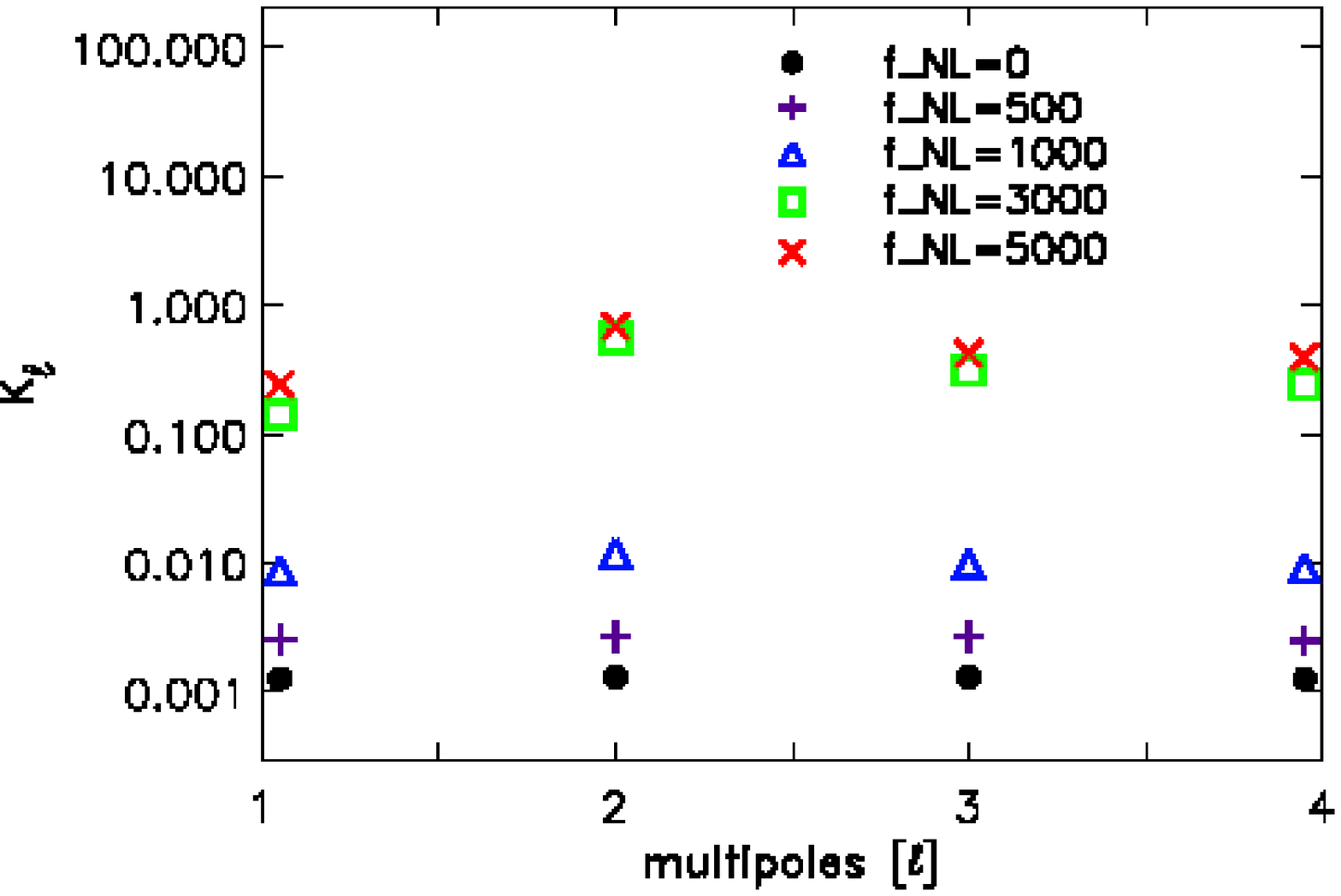}
\hspace{8mm}
\includegraphics[scale=0.45]{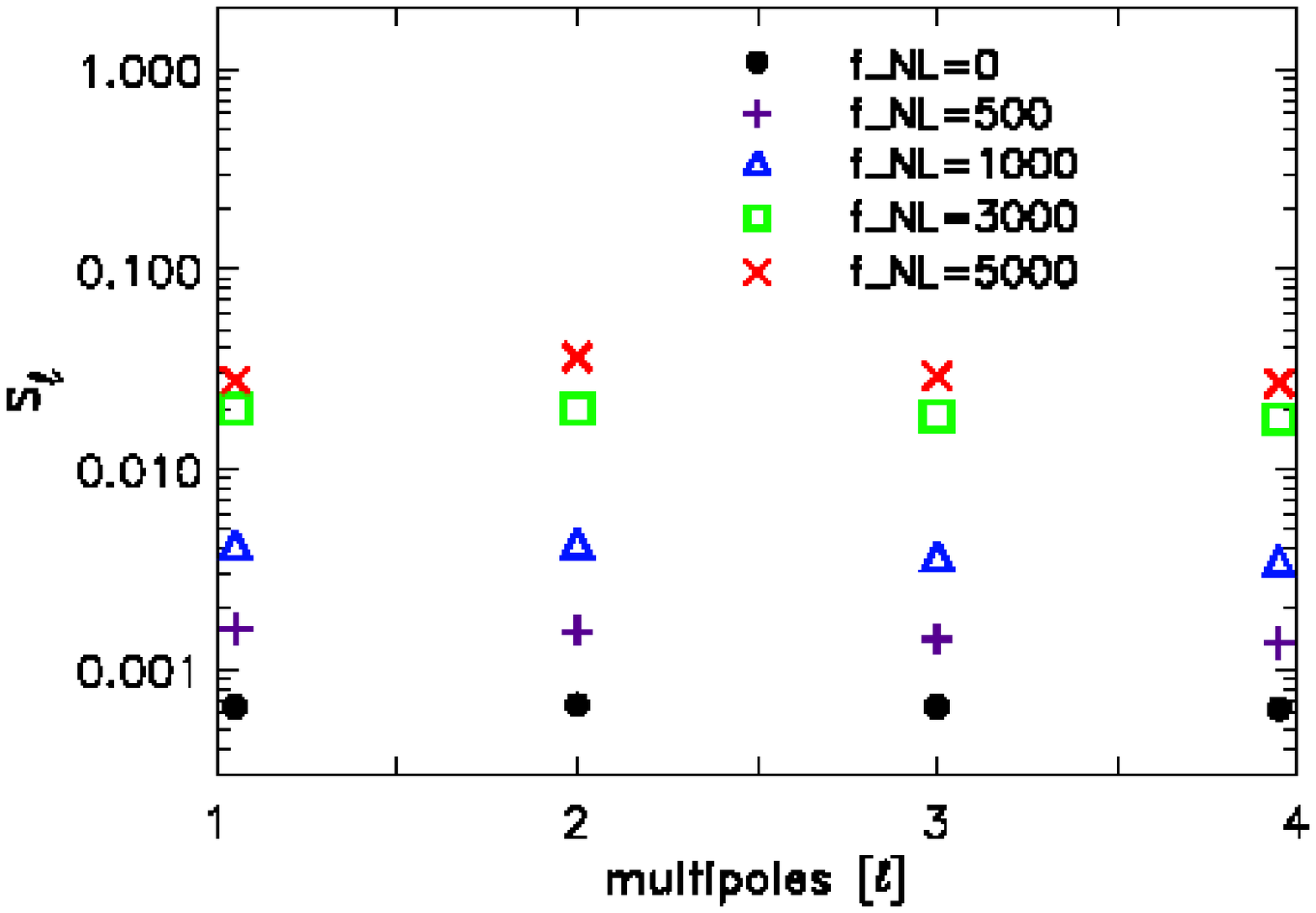}
\caption{Low $ \ell $ angular power spectra, $K_\ell $ and $S_\ell $, of $K$ maps  (left panel) and  of $S$ maps (right panel), respectively, calculated from both Gaussian and non-Gaussian ($ 0 \leq \fNLl \leq 5000  $) simulated CMB maps by using the spherical cells method with $ 48 $ spherical cells.}
\label{Fig-5}
\end{figure*}

Having determined the most appropriated number of cells in the cellular procedure,
we have made an analysis to compare the spherical cells strategy with the spherical
cap procedure employed in Refs.~\cite{BR2009,BR2010,BR2012} (see also related
paper~\cite{Cardona2012}). We have also
examined whether the unexpected \emph{zig-zag} power spectra patterns of $S$ and $K$ maps,
obtained through the spherical caps procedure, is solved by the cells
approach as it is preliminary indicated by Fig.~\ref{Fig-4}.
Table~\ref{Table-2} along with  Fig.~\ref{Fig-5} contain the results of our
calculations carried out by using a total of $24\,000$  $S$ and $K$ simulated maps.
This table shows that both versions of the procedures can be used
to detect \textit{large-angle} ($\ell = 1,2,3,4$) local-type non-Gaussianity
only for fairly large values of the non-linear parameter $f_{\rm NL}^{\rm local}$,
namely $\fNLl \geq 500$ for the  cells procedure, and $\fNLl \geq 3\,000$ for the
spherical caps (here through the $K$ indicator).
Thus, to the extent that for all $\fNLl \simeq 500$, $\chi^2$ test gives greater
values for the cells procedure than for the caps approach, Table~\ref{Table-2}
shows that the spherical cells procedure is more suitable to detect large-angle
($\ell= 1,2,3,4 $) local-type non-Gaussianity than the spherical caps procedure used
in Refs.~\cite{BR2009,BR2010,BR2012}.

The panels of  Fig~\ref{Fig-5} preset the low $\ell$ mean power spectra of the
skewness $S_{\ell}$ (left panel) and kurtosis $K_{\ell}$ (right panel), calculated
from $1\,000$  Gaussian ($f_{\rm NL}^{\rm local}= 0$) maps, and $4\,000$ non-Gaussian
input simulated CMB maps equipped with Non-Gaussianity of the local type for which
$f_{\rm NL}^{\rm local}=500, \,1\,000, $ \\ $ \,3\,000, \,5\,000$.  These panels
reassert the results preliminary detected in Fig.~\ref{Fig-4}, namely they
show that the oscillations in the values of the even and odd modes in the power spectra
$S_{\ell}$ and $K_{\ell}$  that occur in the overlapping spherical caps procedure
(see, for example, Fig.~6 in Ref.~\cite{BR2009}, Fig.~3 in Ref.~\cite{BR2010} and
Fig.~4 of Ref.~\cite{BR2012}) do not come out  in the Fig~\ref{Fig-5}, obtained
through the  spherical cells approach, making clear that this unexpected power spectra
pattern does not have a physical origin, but  rather arises presumably from
the overlapping of the spherical caps.

\section{Concluding Remarks} \label{Sec-5}

The physics of the very early universe can be probed by measuring the statistical
properties of the temperature fluctuations in the CMB data. Since different classes of
such models predict different types and levels of non-Gaussianity for
CMB anisotropies, by studying primordial non-Gaussianity of CMB data one could discriminate
or even rule out either inflationary models or alternative scenarios to the
inflationary paradigm.
In this context, it is important to test CMB data for deviations from Gaussianity by using
different statistical tools to assess and constrain the amount of any non-Gaussian signals
in the data, and extract information about their potential origins. Furthermore,
one does not expect that a single statistical estimator to be sensitive to all possible forms
of non-Gaussianity, which could be present in CMB data.

Recently, two large-angle non-Gaussianity indicators $S$ and $K$ based on skewness and kurtosis
of spherical caps of CMB sky-sphere have been proposed and used to study large-angle
deviation from Gaussianity in masked frequency bands and foreground-reduced full sky CMB
maps~\cite{BR2009,BR2010}.  Even though these indicators can be used to detect non-Gaussianity
signals in CMB data, it has been shown~\cite{BR2012} by using simulated maps that they are
not suitable to discover local non-Gaussianity for the amplitude parameter $\fNLl$  within
the current observational bounds. Moreover, they have found that the power spectra of
$S$ and $K$ maps exhibit an unexpected oscillation in the values of the even and
odd modes in the $ S_ \ell$ and $ K_\ell $ mean spectra, which could have been induced by
overlapping of the patches in the spherical cap procedure.

In this paper we have addressed three interrelated questions regarding advances
of the spherical patches procedures for the detection of large-angle non-Gaussianity.
First, we have examined whether a change in the choice of the
patches would enhance the sensitivity of a new constructive procedure well
enough to detect large-angle non-Gaussianity of local type with $f_{\rm NL}^{\rm local}$
within the Planck bounds. Second, we have studied whether this new procedure
with suitable non-overlapping choice of spherical patches
would be capable to smooth out the undesirable oscillation pattern in
the power spectra of the associated skewness and kurtosis maps.
Third, we have additionally made a comparative analysis of the new spherical cells procedure
with the caps routine of ~\cite{BR2009,BR2010,BR2012} to find out their  relative
strength and weakness in the study of large-angle ($\ell =1, 2, 3, 4$) non-Gaussianity.

To this end, we have employed the two procedures along with $9\,000$
simulated  CMB temperature maps equipped with non-Gaussianity of local type with
various amplitudes. From these simulated maps, which include
the Gaussian ones with $f_{\rm NL}^{\rm local}=0$, we have generated $24\,000$
$S$-maps and $24\,000$ $K$-maps (for $48$ and $192$ cells and $3\,072$ caps),
calculated the low $\ell$ mean power
spectra $S_\ell$ and $K_\ell$, made a study of the sensitivity and strength,
and determined the limitations of non-Gaussian estimators $S$ and $K$.
The results of our analyses  are summarized in Tables~\ref{Table-1} and
\ref{Table-2} together with Fig.~\ref{skewness map} to Fig.~\ref{Fig-5}

The negligible value of $\chi^2$ analysis collected together in Table~\ref{Table-1}
shows, on the one hand, that the spherical cells new procedure, with both $48$ and $192$ cells,
do not have sufficient sensitivity to detect local non-Gaussianity for the amplitude
parameter within the Planck bounds $20 \leq  f_{\rm NL}^{\rm local} \leq 56$.
Figure ~\ref{Fig-4} complements this result by showing the nearly overlapping  of the
symbols for $\fNLl=0, 20, 56$. On the other hand, Table~\ref{Table-1} also makes clear
that for non-Gaussinity amplitude $\fNLl \geq 500$ the spherical cells procedure
with $48$ cells is more suitable to detect local non-Guassinity than with $192$ cells.

Table~\ref{Table-2} shows that both versions of the procedures can be used
to detect large-angle ($\ell = 1,2,3,4$) local-type non-Gaussianity
for reasonably large values of the amplitude parameter $f_{\rm NL}^{\rm local}$, namely
$\fNLl \geq 500$ for the  cells procedure, and $\fNLl \geq 3\,000$ for the spherical caps.
Thus, Table~\ref{Table-2} shows that the spherical cells procedure is more suitable
to detect large-angle local-type non-Gaussianity than the spherical caps procedure
of Refs.~\cite{BR2009,BR2010,BR2012}.

Finally, the panels of Fig~\ref{Fig-4} and Fig.~\ref{Fig-5}
show that  the oscillations in the values of the even and odd modes in the
power spectra $S_{\ell}$ and $K_{\ell}$ of the overlapping spherical caps procedure
do not come about  in these figures obtained through the  spherical cells
procedures. This makes clear that the power spectra unexpected patterns of the
caps procedure do not have a physical origin, but  rather  arises presumably from
the overlapping of the spherical caps.

\begin{acknowledgements}
We would like to thank Martin Kunz for helpful discussions, important comments and suggestions.
We are grateful to A.F.F. Teixeira for reading the manuscript and indicating some omissions and typos.
M.J. Rebou\c{c}as acknowledges the support of FAPERJ under a CNE E-26/102.328/2013 grant. A. Bernui,  and M.J. Rebou\c{c}as thank  CNPq for the grants under which this work was carried out. W. Cardona is supported by COLCIENCIAS and acknowledges financial support by CNPq in the early stage of this work.  Some of the results in this paper were derived using the HEALPix package\cite{Gorski05}.
\end{acknowledgements}



\end{document}